\documentclass{eas}
\usepackage{graphicx}
\usepackage{epsfig}
\usepackage{overrulehere}

\begin{document}

%%-----------------------------
%%      the top matter
%%-----------------------------
\title{Earthshine observations of an inhabited planet}
\author{Enric Pall\'e}\address{Instituto de Astrofisica de Canarias, Via Lactea s/n, 38205 La Laguna, Tenerife, Spain}
%\author{...}\address{...}
%\author{...}\address{...}
%
%
\begin{abstract}

Earthshine is sunlight that has been reflected from the dayside Earth onto the dark side of the Moon and back again
to Earth. In recent times, there has been  renewed interest in ground-based visible and near-infrared measurements
of  earthshine as a proxy for exoplanet observations. Observations of  earthshine allow us to explore and
characterize the globally integrated photometric, spectral and polarimetric
features of the Earth, and to extract precise information on  the distinctive characteristics of our planet,
and life in particular. They also allow us to quantify how this feature changes with time and orbital configuration.
Here we present a brief review of the main earthshine observations and results.

\end{abstract}
\maketitle
%%-----------------------------
%%      your text
%%-----------------------------
\section{Introduction}

Over the past few years, we have developed the capacity to discover planets orbiting around stars other than the Sun,
and the number of detections is increasing exponentially. Even though we are not yet capable of detecting and exploring
Earthlike planets, ambitious missions are already being planned for the coming decades. In the near future, it is likely
that Earth-size planets will be discovered, and efforts will then be directed towards
obtaining images and spectra from them.

The resulting spatially integrated observations will need to be compared to observations of similar resolution of the
only planet that we know to be inhabited, namely Earth. The scientific and technological advances of the XXth century
have brought a finer and finer detailed knowledge of our planet. Routinely, plankton blooms, city night lights or crop
health indices are measured from space. However, when observing an extrasolar planet, all the reflected starlight and the
radiated emission from its surface and atmosphere will be integrated into a single point, and most of these features will
disappear in the noise when global averages are taken.

Despite the fact that Earth is our own planet, observations from a remote perspective are limited. They can only be
obtained from a very remote observing platform, from a complex calibration of low Earth orbit satellite observations,
or from the ground by observing the earthshine reflected on the dark side of the Moon.

\begin{figure}
\begin{center}
\epsfig{file=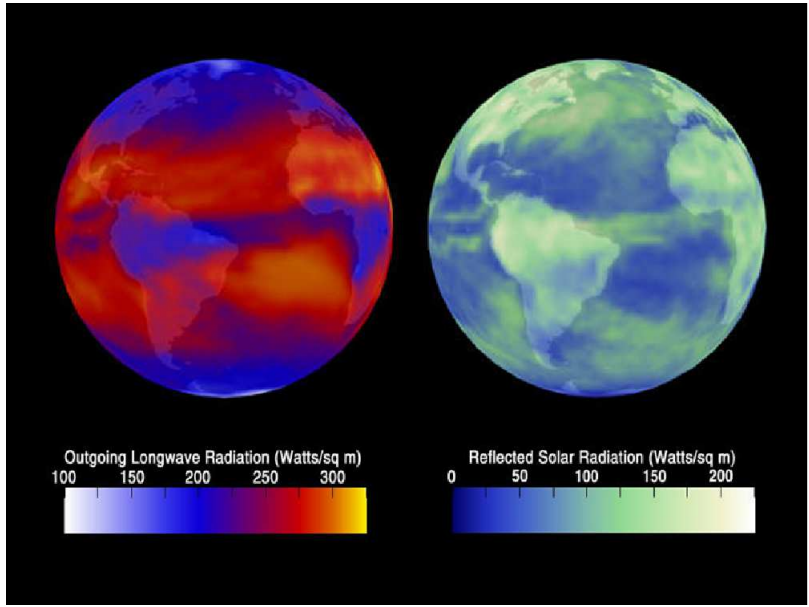,height=8.0cm} \caption{The spatial distribution of the thermal radiation emitted into
space from the Earth's surface and atmosphere (left), and the sunlight reflected  back into space by oceans, land, aerosols
and clouds (right). Image credit: NASA} \label{maps2}
\end{center}
\end{figure}

What these observations reveal about our planet will be highly dependent on several factors. The observing geometry
will determine whether one samples either the full surface of the planet or just one of the hemispheres. The spectral
range is also important, as the light reflected or emitted by Earth is not uniformly distributed (see Figure~\ref{maps2}).
Finally, timing is also important. Our planet has undergone evolution with time, so an external observer would obtain a
different view of Earth if observing it at different epochs (Kaltenegger {\em et al.} 2007). In this paper we
 concentrate on modern Earth observations, mostly in the visible range, as obtained from earthshine observations.

\section{The Earthshine}

Earthshine is sunlight reflected from the dayside Earth onto the dark side of the Moon and reflected back again to Earth.
The term ``dark side'' refers to the portion of the lunar surface that, at any instant, faces the Earth but not the Sun.
Both earthshine, from the dark side of the moon, and moonlight, from the bright side of the moon, are transmitted through
the same airmass just prior to detection and thus suffer the same extinction and imposed absorption features. Their ratio
is the averaged reflection coefficient (or albedo) of the global atmosphere, the global atmosphere being defined as the
portion of  dayside Earth simultaneously visible from the Sun and the Moon.

Already in the XVIth century, Leonardo da Vinci had correctly deduced the nature of the Earthshine. In his {\itshape Codex
Leicester} (circa 1510), he states his belief that the Moon possessed an atmosphere and oceans, and that it was a fine
reflector of light because it was covered with so much water. He also speculated about how storms on Earth could cause
the Earthshine to become brighter or dimmer, which is indeed observable with modern instrumentation. After Leonardo,
others (such as Galileo) continued to observe the earthshine in historical times. In recent times, however, there has been
a renewed interest in ground-based visible and near-infrared measurements of the earthshine as a proxy for exoplanet
observations (Woolf {\em et al.} 2002), and also for climate studies (Pall\'e {\em et al.} 2004).

\begin{figure}
\begin{center}
\epsfig{file=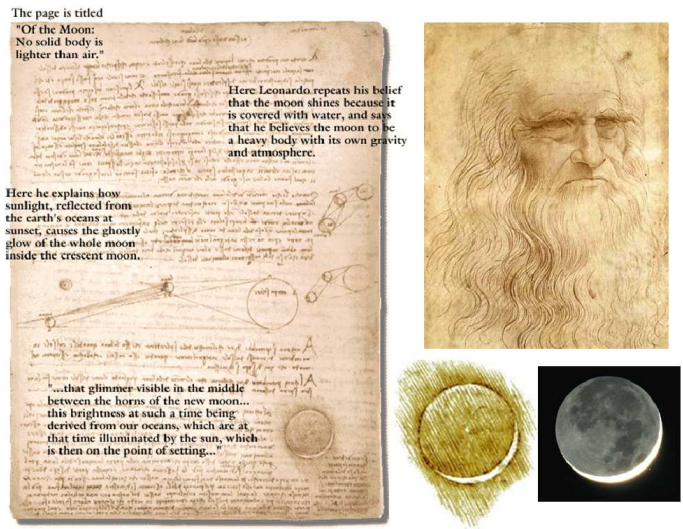,height=7.0cm}
\caption{Reproduction of a page of Leonardo da Vinci's {\it Codex Leicester}. Also shown is a image of da Vinci in his
{\it Autoportrait} and a comparison of a real picture of the earthshine with one of his drawings. Image credit:
American Museum of Natural History Library} \label{codex}
\end{center}
\end{figure}

\section{Earth's reflected light}

\begin{figure}
\begin{center}
\epsfig{file=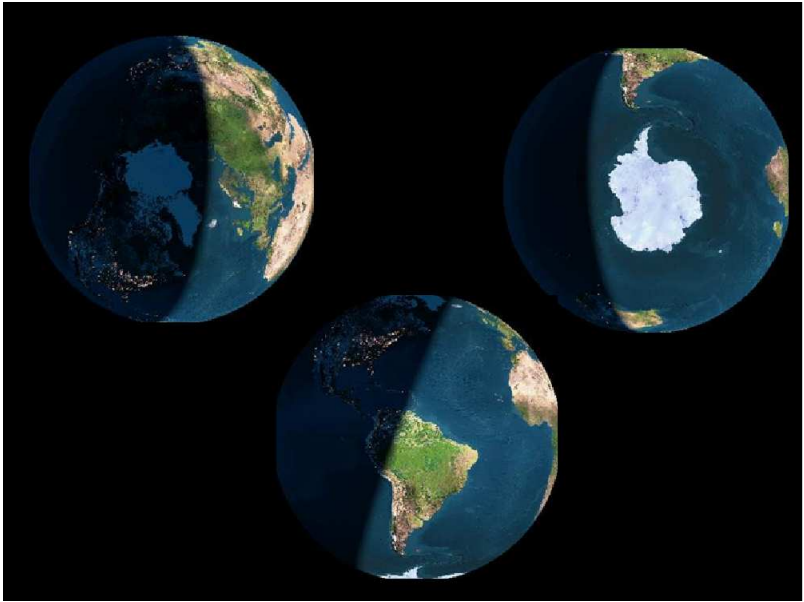,height=8.0cm} \caption{The Earth from different viewpoints. The three images show the
Earth for the exact same date and time (2003/11/19 at 10:00 UT) but from three different perspectives: from $90^\circ$ above
the ecliptic (north polar view) (right), from $90^\circ$ below the ecliptic (south polar view) (left),  and from within the
ecliptic plane (centre).}
\label{views}
\end{center}
\end{figure}

The light reflected by the Earth in the direction of a hypothetical distant observer will change in time, depending
on the orbital phase, rotation, seasonality (tilt angle) and weather patterns. Each observing perspective of Earth will have
its unique features and will present different photometric variability. Figure~\ref{views} illustrates this point: Several
views of Earth are represented for the exact same date and time, but judging from the visible scenery features, the three
images could well represent three different planets. Note, however, that the figures are misleading in the sense that clouds are missing.

A number of earthshine measurements have documented changes with phase and rotation but  are necessarily limited in their
range of observations.  From these measurements the photometric variability of the Earth is found to be of the order of 10--15\%
from night to night  during one Earth rotation (Qiu {\em et al.} 2003).
For an observer outside the solar system, Earth observations will differ from the mostly equatorial perspective offered by the
earthshine measurements, or the global averages obtained from polar orbiter satellites. The total reflected flux in a given direction,
$\beta$, can be calculate using

\begin{equation}
F_{e}(\beta) = S \pi R_{e}^{2}  p_{e}f_{e}(\beta), \label{Fe}
\end{equation}
where $S$ is the solar flux at the top of the Earth's atmosphere (1370 $W/m^{2}$), and $R_{e}$, $p_{e}$, and $f_{e}(\beta)$ are
the radius, geometric albedo and phase function of the Earth, respectively. There is a systematic variation of $p_{e}f_{e}(\beta)$
throughout the Earth's orbital period (sidereal year), and fluctuations of $p_{e}f_{e}(\beta)$ about its systematic behaviour are
caused by varying terrestrial conditions, including weather and seasons (Pall\'e {\em et al.} 2004).

Among other physical properties, the identification of the rotation rate of an exoplanet with relative accuracy will be important
for several reasons. If the rotation period of an Earthlike planet can be determined accurately, one can then fold the photometric
light curves at the rotation period to study regional properties of the planet's surface and/or atmosphere. With phased light curves
it could be possible study local surface or atmospheric properties with follow-up photometry, spectroscopy and polarimetry to detect
surface and atmospheric inhomogeneities and to improve the sensitivity to localized biomarkers. Exoplanets, however, are expected to
deviate widely in their physical characteristics: if they have no strong surface features (Mercury or Mars), or they are completely
covered by clouds (Venus and the giant planets), determining the rotational period may be an impossible task.

Pall\'e {\em et al.} (2008) found that scattered light observations of the Earth could accurately identify the rotation period of the
Earth's surface. This is because large-scale time averaged cloud patterns are tied to the surface features of Earth, such as continents
and ocean currents. This relatively fixed nature of clouds is the key point that would allow Earth's rotation period to be determined
from afar. The lifetime of large-scale cloud systems on Earth is typically about one to two weeks (roughly 10 times the rotational
period). In fact, Earth may well be the only one of the major planets for which a rotational period can be easily established from a
distance of several AU.

\subsection{Terrestrial clouds}

Clouds are common on  solar system planets, and even on satellites with dense atmospheres. On Earth, clouds are continuously
forming and disappearing, covering an average of about 60\% of the Earth's surface (Rossow {\em et al.} 1996). This
feature is unique in the solar system to Earth: only the Earth has large-scale cloud patterns that partially cover the
planet and partly reveal the rocky surface, and change on timescales comparable to the rotational period. This is because
the temperature and pressure on the Earth's surface allow for water to change phase with relative ease from solid to liquid to gas.

\begin{figure}[H]
\begin{center}
\epsfig{file=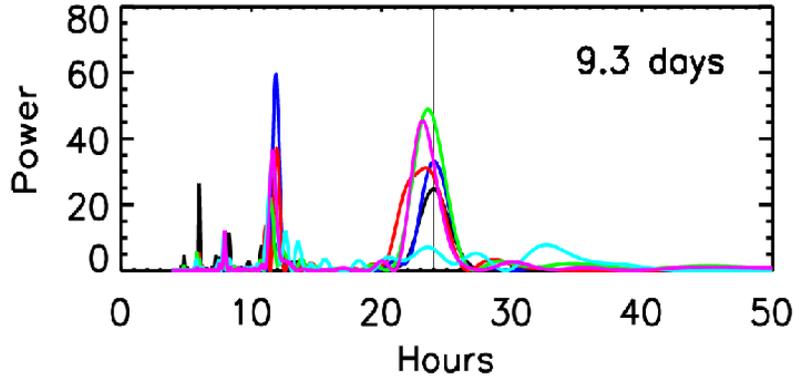,height=6.0cm}
\caption{Periodogram analysis of the Earth's scattered light from an equatorial viewing angle. Two months of
reflected light simulations are subdivided in six equally long time series, and a periodogram analysis is performed on each series.
In the figure, different colours indicate different data subperiods. Note the appreciable decrease in the retrieved rotation rate for
some of the time series. Other peaks in the periodogram indicate shorter periodicities introduced by the Earth's land/ocean
distribution. Adapted from Pall\'e {\em et al.} (2008).} \label{shift1}
\end{center}
\end{figure}

Pall\'e {\em et al.} (2008) studied the periodicity shifts that appear in modelled Earth observations (see Figure~\ref{shift1}),
and concluded that they are introduced by the large-scale wind and cloud patterns. As these patterns change, the apparent
rotational period of the Earth is sometimes shorter than 24 hours. Thus, for an extrasolar Earth-like planet, photometric
observations could be used to infer the presence of a ``variable'' surface (i.e.\ clouds), even in the absence of spectroscopic
data.  This would strongly suggest the presence of liquid water on the planet's surface, especially if the effective temperature
of the planet was also determined by other means.

\subsection{Glint}

Although the reflectivity of water is very low at high and medium angles of incident light, it increases tremendously at small
angles of incident sunlight. This is observable from space on the sun-illuminated side of the Earth near the terminator, and it
it known as the glint (Campbell {\em et al.} 2003). Waviness, however, causes an appreciable reduction in the glint magnitude.

\begin{figure}
\begin{center}
\epsfig{file=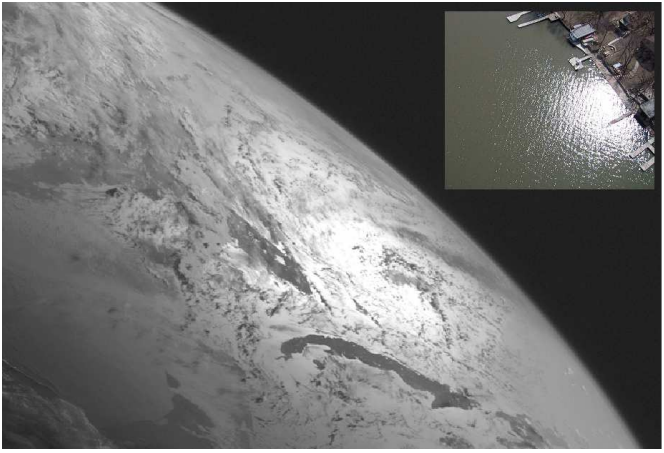,height=6.0cm}
\caption{Sun-glint in the Sargasso Sea, obtained by GOES-West, on 2000 June 22. The top corner illustrates the glint phenomenon
at a smaller scale. Image credit: NASA} \label{glint}
\end{center}
\end{figure}

To assess the detectability of specular reflection from an ocean in the (disk-averaged) reflected light from a distant planet,
Williams \& Gaidos (2008) used a model to simulate the orbital brightness variation, or ``light curve'', of Earth. They concluded
that ocean-bearing planets with edge-on orbital geometries will exhibit large changes in their apparent reflectivity due to
specular reflection. Similar phase brightening of the crescent Earth have been empirically identified in earthshine data
(Pall\'e {\em et al.} 2003). Earthshine has also been found to be strongly polarized, which is further indication of specular
reflection of sunlight off the oceans (Stam {\em et al.} 2006).

\section{Earth's spectral signatures}

Hitchcock \& Lovelock (1967) pointed out that the remote detection of life forms might be possible by studying the atmospheric
composition of a planet. On Earth, the presence of $H_{2}O$, $O_{2}$ and $O_{3}$ in its atmosphere is clearly marked in the
spectra. This particular combination of atmospheric constituents, because they are in disequilibrium, strongly signal  the
presence of life, the driver of the disequilibrium.

In the past, there have been numerous attempts to observe the spectrum of the pale blued dot from a distance in different
spectral regions. Among other efforts, observations of the Earth as a planet from afar have been made in the visible range from
the Galileo spacecraft (Sagan {\em et al.} 1993).

With the aim of improving the spectral resolution and the sampling of seasonality and phase changes, spectroscopic measurements
of the earthshine have also been taken from ground-based observatories and analysed at visible wavelengths  (Woolf {\em et al.}
2002); Monta\~n\'es-Rodr\'iguez {\em et al.} 2005 and 2006), and in the near infrared wavelengths (Turnbull {\em et al.} 2006).

\begin{figure}
\begin{center}
\epsfig{file=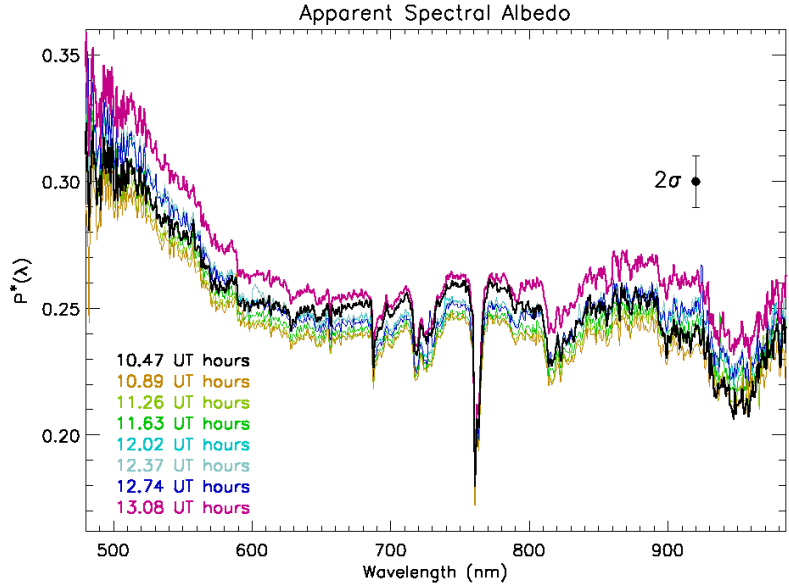,height=8.0cm} \caption{Several earthshine spectra, taken with the 60$''$ Telescope at Palomar
Observatory in November 2003. Different colours indicate different observing times during the night. The main features of the
Earth's reflectance in this region include an enhancement due to  Rayleigh scattering in the blue part of the Chappuis
$O_3$ band, which contributes to the drop above $500$ nm. Atmospheric absorption
bands due to oxygen, the sharpest, $A-O_2$ at $760$ nm, and water vapour are clearly detected. Adapted from Monta\~n\'es-Rodr\'iguez
{\em et al.} (2005)} \label{visspec}
\end{center}
\end{figure}

The overall shape of the Earth's spectrum in the visible region shows interesting and peculiar signatures. The most prominent
is the Rayleigh scattering, the main source of opacity in the Earth's atmosphere, which shows an enhancement in reflectivity
towards the blue part of the spectra (Figure~\ref{visspec}). The strong wavelength dependence of the scattering ($\lambda^{-4}$)
means that blue light is scattered much more than red light. The Rayleigh scattering of sunlight by the Earth's atmosphere is the
main reason why the sky is blue, and our planet is known as the blue planet. Except for Neptune and Uranus, no other solar system body
shows this strong Rayleigh feature in the blue. However, Neptune and Uranus' predominant blue colour does not come from Rayleigh
scattered light, but from the absorption of red and infrared light by methane gas in their atmospheres.

At shorter wavelengths (${\rm <310\,\,nm}$) ozone absorption dominates over the Rayleigh scattering, causing a strong decrease
in reflectance, which has also been measured through earthshine observations (Hamdani {\em et al.} 2006). In the near-infrared, the
 Earth's spectra is dominated by strong absorption bands of $H_{2}O$, $CO_{2}$ (Turnbull {\em et al.} 2006).

Indications of a sharp increase in reflectivity near $0.720 \mu m$ have also been identified as due to reflectance from vegetation
(Monta\~n\'es-Rodr\'iguez {\em et al.} 2006). However, this would be an ambiguous biosignature on an exoplanet, because vegetation
is not the only possible source of the reflectivity bump
(Tinetti {\em et al.} 2006).

\begin{figure}
\begin{center}
\epsfig{file=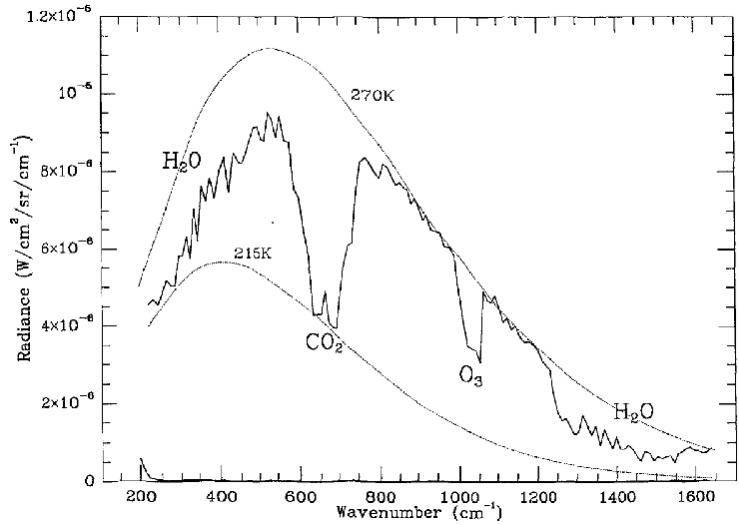,height=7.0cm}
\caption{Normalized calibrated spectral radiance of the Earth (central heavy curve).
The data have been normalized to the highest signal. The noise equivalent spectral
radiance is the lower heavy curve near the $x$ axis. Dotted curves sow blackbodies
at 215 and 270 K. Adapted from Christensen \& Pearl (1997)} \label{IRdata}
\end{center}
\end{figure}

In the infrared range, earthshine observations cannot be obtained using the earthshine due to the strong absorption of the Earth's
atmosphere and the emission from the Moon. Thus, observations of the infrared spectrum of Earth are only available from occasional
remote spacecraft observations. The thermal emission spectrometer (TES) on the Mars Global Surveyor spacecraft acquired one  such
observation of the Earth from a distance of 4.7 million km on 1996 November 24 (Christensen \& Pearl 1997). The thermal emission
of the Earth dominates the IR spectrum, corresponding to its effective temperature of 288 K (see Figure~\ref{IRdata}). The IR
brightness peaks around 10 $\mu$m and then decays slowly. Spectral features of the gases carbon dioxide ($CO_2$), water vapour ($H_{2}O$),
and ozone ($O_3$) dominate the Earth's spectra in this range, as well as methane ($CH_4$) and several other minor constituents. Radiation
at the centre of the $CO_2$ band arises mainly from the lower stratosphere; near 650 and 700 $cm^{-1}$ from near the tropopause; and
further into the band wings from the troposphere and surface. Thus, in the disc-averaged sense, the spectrum indicates a warm stratosphere
above a tropopause somewhat colder than 215 $K$.

\section{Earthshine polarization measurements}

The light emitted by the sun is unpolarized; however, when it interacts with the Earth's atmosphere it becomes polarized by
 transmission, reflection, refraction, or scattering (Hecht \& Zajac 1997). The degree of polarization of the globally integrated
light scattered from Earth depends on the cloud coverage, together with the degree of polarization of each of the exposed areas of its surface.

Dollfus (1957) was the first to take observations of  earthshine polarization and compare them to direct measurement of polarization
from the ground from balloon observations. From his measurements, he concluded that the atmospheric polarization is  far larger than
that introduced from the ground and noted that `for an extra-terrestrial observer it would be diluted by the intense, but little
polarized, light from the background.' He also noted how the degree of polarization was phase dependent. More recently, based on
polarization observations of the Earth made from the Polder satellite, Wolstencroft \& Breon (2005) determined that the degree of
linear polarization, at 443 nm and $90^{o}$ scattering angle, is 23\% for averaged cloud cover conditions.

\begin{figure}
\begin{center}
\epsfig{file=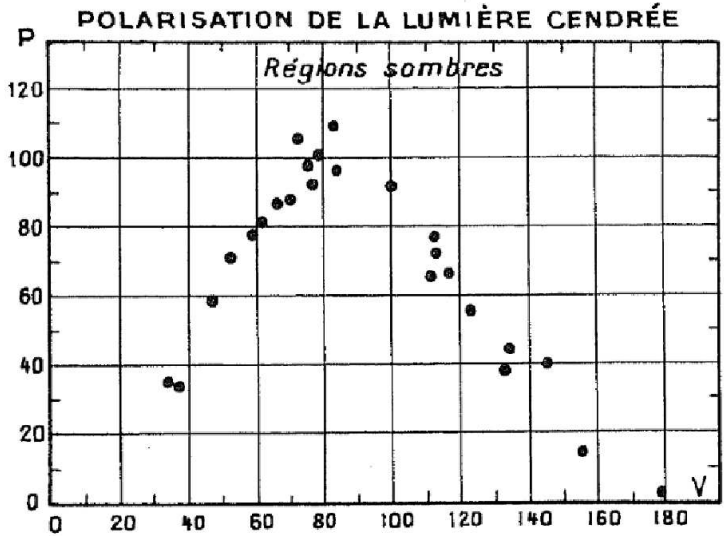,height=8.0cm} \caption{The degree of polarization of the earthshine, plotted as a function of phase.
These are the only existing globally-integrated measurements of the Earth's polarization, and they are only indicative, because
reflection from the Moon introduces a variable depolarization factor. Adapted from Dollfus (1957). } \label{dollf}
\end{center}
\end{figure}

The rotation of a planet with surface features such as continents and oceans should modulate the polarized reflectances in a simple and
predictable manner. The models (McCullough 2008) predict the shape of the phase function of the Earthshine's linear polarization as
observed by Dollfus (see Figure~\ref{dollf}). The maximum polarization agrees with Dollfus's observations but is approximately twice as
large as that predicted by Wolstencroft \& Breon (2005). Thus, linear polarization could be a potentially useful signature of oceans and
atmospheres of Earth-like extrasolar planets.

Biotic material, with its helical molecular structure, is known to produce circular polarization of reflected light in the visible
range (Pospergelis 1969; Wolstencroft \& Raven 2002). Observed on the Moon, chiral signatures in the Earthshine are expected of the
order of $V/I$ $10^{-4 ...-5}$, assuming a dilution of the circular polarimetric signal caused by vegetation due to partial cloud
coverage (factor 10), and by depolarization on the lunar surface up to a factor 10 (DeBoo {\em et al.} 2005). Preliminary measurement
of non-zero Stokes $V/I$ during a specific phase of the Earthshine, were observed by Sterzik {\em et al.} (2008) but could not
unambiguously be interpreted as the signature of biotic homochirality, and therefore life, on Earth.

On Earth, water surfaces provide the only significant sources of thermal infrared polarized radiation, while emission from the atmosphere
and ground is almost always unpolarized to any practical degree (Shaw 2002). Thus, the degree of polarization of the infrared flux emitted
from Earth, observed as a planet, is negligible.

\section{ET's Conclusions}

In conclusion, if an extraterrestrial (or extrasolar) observer was looking at the Earth from an astronomical distance, and similarly
if we were observing an Earth-like exoplanet, we would be able to determine some of its properties. From al these properties, those
regarding the possibility of finding life would be:

\begin{itemize}
\item{Habitability signatures:}
\subitem{Atmospheric composition}
\subitem{Variability and albedo}
\subitem{Spectral shape (Rayleigh scattering, atmospheric properties)}
\subitem{Rotational period}
\subitem{Presence of oceans (glint)}
\subitem{Presence of dynamic weather ... and possibility of liquid water}

\item{Life signatures:}
\subitem{Atmospheric disequilibrium caused by life}

\item{Complex life (plant/animal) signatures:}
\subitem{Presence of complex life (red or other edges)}
\subitem{Fluorescence}
\subitem{Presence of biomass from circular polarization}

\item{Intelligent life signatures:}
\subitem{Direct emissions: Radio, lasers, etc.}
\subitem{Indirect emissions: Night lights, CFCs, ...}
\subitem{Unexplainable atmospheric composition /properties}
\end{itemize}

Of all the signatures described above, only the habitability and life signatures are likely to be observed in extrasolar planets
during the coming decades. Even then, it will require instruments and observing techniques which are nowadays only in the conceptual
phase. It is extremely unlikely that we will detect any of the complex life signatures described in this paper, with the possible
exception of the red edge, considering that exoplanets may exist in a wide range of physical properties. Even then, however, it will
be very difficult to attribute the red edge to biota. The signature of intelligent life, although easier to detect from the technical
point of view, requires that the alien civilization was trying to contact us, thus the author considers them very unlikely to be detected
 in the near future, but with an extreme desire of being proven wrong.

{\bf Acknowledgements}

The contents of this paper are examined in much greater detail in the forthcoming book {\it The Earth as a Planet}, by M. Vazquez,
E. Pall\'e  and P. Monta{\~n}{\'e}s Rodriguez.

%%-----------------------------
%%      your bibliography
%%-----------------------------


\begin{thebibliography}{99}




\bibitem[]{} Campbell, D.B. \etal\ 2003, Sci, 302, 431-434
\bibitem[]{} Christensen, P.R. \& Pearl, J.C. 1997, JGR, 102, 10875-10880
\bibitem[]{} Deboo, B.J. \etal\ 2005, 44, 5434-5445
\bibitem[]{} Dollfus, A. 1957, Suppl. Ann. Astroph. 4, 3-114
\bibitem[]{} Hamdani, S. \etal\ 2006, A\&A, 460, 617-624
\bibitem[]{} Hetch, E. \& Zajac, A. 1997, Optics, Addison Wesley Publishing Company; 3rd edition
\bibitem[]{} Hitchcock, D.R. \& Lovelock, J.E. 1967, Icarus, 7, 149-159
\bibitem[]{} Kaltenegger, L. \etal\ 2007, ApJ, 658, 598-616
\bibitem[]{} McCullough, B. 2008, ApJ (submitted)
\bibitem[]{} Monta\~n\'es-Rodr\'iguez \etal\ 2005, ApJ, 629, 1175-1182
\bibitem[]{} Monta\~n\'es-Rodr\'iguez \etal\ 2006, ApJ, 651, 544-552
\bibitem[]{} Pall\'e, E. \etal\ 2003, JGR, 108, 4710
\bibitem[]{} Pall\'e, E.  \etal\ 2004, Sci, 304, 1299-1301
\bibitem[]{} Pall\'e, E. \etal\ 2008, ApJ, 676, 1319-1329
\bibitem[]{} Pospergelis, M.M. 1969, SvA, 12, 973
\bibitem[]{} Qiu, J. \etal\ 2003, JGR, 108, 4709
\bibitem[]{} Rossow, W.B. \etal\ 1996, WMO/TD-No. 737, WMO, Geneva, 115pp.
\bibitem[]{} Sagan, C.  \etal\ 1993, Nat, 365, 715
\bibitem[]{} Shaw, J.A. 2002, SPIE, 4819, 129-138
\bibitem[]{} Stam, D.M. \etal\ 2006, A\&A,452,669-683
\bibitem[]{} Sterzik, M. 2008, (in preparation)
\bibitem[]{} Tinetti, G. \etal\ 2006, Astrobiology, 6, 881-900
\bibitem[]{} Turnbull, M.C. \etal\ 2006, ApJ, 644, 551-559
\bibitem[]{} Williams, D.M. \& Gaidos, E. 2008, ArXiv e-prints, 0801.1852
\bibitem[]{} Wolstencroft, R.D. \& Breon, F.M. 2005, PASPC, 343, 211
\bibitem[]{} Wolstencroft, R.D. \& Raven, J.A. 2002, Icarus, 157, 535-548
\bibitem[]{} Woolf, N.~J. \etal\ 2002, ApJ, 574, 430-433
















\end{thebibliography}
\end{document}